\documentstyle[epsf,12pt]{article}
\def\doublespace{\lineskip      .25 ex
\baselineskip 5.0 ex
\lineskiplimit 0 ex
\parskip 1.0 ex plus.50 ex minus .25 ex}%
\begin{document}
\newcommand{\be}{\begin{equation}}
\newcommand{\ee}{\end{equation}}
\doublespace
\centerline{\bf EVOLUTION OF RELATIVISTIC POLYTROPES}
\vspace*{0.035truein}
\centerline{\bf IN THE POST--QUASI--STATIC REGIME}
\vspace*{0.37truein}
\centerline{\footnotesize {\large L. Herrera
\footnote{ Escuela de F\'{\i}sica, Facultad de Ciencias, Universidad Central de
Venezuela, Caracas, Venezuela.}
\footnote{Postal address:
Apartado Postal 80793, Caracas 1080A, Venezuela. e--mail:
laherrera@telcel.net.ve}
and W. Barreto}
\footnote{Grupo de  F\'\i sica Te\'orica, Departamento de F\'\i sica,
 Facultad de Ciencias, Universidad de los Andes,  M\'erida, Venezuela.
e--mail: wbarreto@ula.ve}}
\baselineskip=12pt
\vspace*{0.21truein}
\date{\today}
\begin{abstract}
\doublespace
A  recently presented method for the study of evolving
self-gravitating relativistic spheres is applied to the description of the
evolution of relativistic polytropes. Two different definitions of
relativistic polytrope, yielding the same
Newtonian limit, are considered. Some examples are explicitly worked out,
describing collapsing polytropes and bringing out the differences between
both types of polytropes.

\vspace*{0.21truein}
\end{abstract}
\newpage
\doublespace
\section{Introduction}
The use of polytropic equations of state in the study of the stellar
structure has long and a venerable history \cite{Chandra,Shapiro,Weigert}
(and references
therein). Its great sucess stems, mainly,  from its simplicity and from the
fact that it can be used to describe a large number of different
situations.

It is therefore not surprising that a great deal of work  has been devoted
to the study of polytropes in the context of general relativity 
\cite{Tooper,Bludman,Ugla,Jap} (and references
therein). Nevertheless, since the Lane--Emden equation, which is the
cornerstone in the study of polytropic spheres,
is based on the assumption of hydrostatic equlibrium, almost  all works
done so far (to our knowledge) on polytropic equations of state, concern
spheres in hydrostatic equilibrium
(collapsing ``Newtonian'' polytropes with
$n=3$, have been considered by Goldreich and Weber \cite{Goldreich}).

However, during their evolution, self--gravitating objects may pass
through phases of intense dynamical activity, with time scales of the order
of magnitude of (or even smaller than) the hydrostatic time scale, and  for
which the static (or the quasi--static)
approximation is clearly not reliable (e.g. the collapse of very massive
stars \cite{Iben} and the quick collapse phase
preceding neutron star formation \cite{myra} (and
references therein). In these cases it is mandatory to take into account
terms which describe departure from
equilibrium. Accordingly, it is our purpose in this work to study the
evolution of polytropes out of hydrostatic equilibrium.

For doing so, we shall make use of  an approach for
modeling the evolution of self--gravitating spheres, which does not require
full numerical integration of time dependent Einstein equations
\cite{Herrera}. The motivation for this, is based on the following
argument: It is true that numerical methods \cite{Lehner} (and
references therein)
are enabling researchers to investigate systems which are extremely
difficult to handle analytically. In the case of General Relativity,
numerical models have proved valuable for investigations of strong
field scenarios and have been crucial to reveal unexpected phenomena
\cite{choptuik}. Even specific difficulties associated with numerical
solutions of partial differential equations in presence of shocks are
being overpassed
 \cite{font}. By these days what seems to be the main limitation for
 numerical relativity is the computational demands for 3D evolution,
 prohibitive in some cases \cite{winicour}. Nevertheless, it is obviously
simpler (in general) to solve ordinary than partial differential equations
and furthermore purely
numerical solutions usually hinder to catch general, qualitative, aspects
of the process. Instead, the proposed method, starting from any interior (analytical or
numerical) static
spherically
symmetric (``seed'') solution to Einstein equations, leads to a system of
ordinary differential equations for quantities evaluated at the boundary
surface of the fluid distribution,
whose solution (numerical), allows for modeling the dynamics of
self--gravitating
spheres, whose static limit is the original ``seed'' solution.

The approach is based on the introduction of a set of conveniently defined
``effective'' variables, which are effective pressure and energy density, and an
heuristic ansatzs on the latter
\cite{Herrera}, whose rationale and justification become intelligible
within the context of the post--quasistatic appproximation defined below. In
the quasistatic approximation
(see below), the effective variables coincide with the corresponding
physical variables (pressure and density) and  therefore the method may be
regarded as an iterative method
with each consecutive step corresponding to a stronger departure from
equilibrium.
In  this work we shall restrain ourselves to the post--quasistatic level
(see  section 4 for details).

The fluid distribution under consideration will be assumed to
be dissipative. Indeed, dissipation due to the emission of massless
particles (photons and/or neutrinos) is a characteristic process in the
evolution of massive stars. In fact, it seems that the only plausible
mechanism to carry away the bulk of the binding energy of the collapsing
star, leading to a neutron star or black hole is neutrino emission
\cite{1}. Consequently, in this paper, the matter distribution forming
the selfgravitating object will be described as a dissipative
fluid, which in the equilibrium regime satisfies a polytropic equation of state.

On the other hand, in the treatment of radiative transfer within stellar
objects, two different approximations are usually adopted: difussion and
streaming out.

In the diffusion approximation, it is assumed that the energy flux of
radiation (as that of
thermal conduction) is proportional to the gradient of temperature. This
assumption is in general very sensible, since the mean free path of
particles responsibles for the propagation of energy in stellar
interiors is in general very small as compared with the typical
length of the object.
Thus, for a main sequence star as the sun, the mean free path of
photons at the centre, is of the order of $2\, cm$. Also, the
mean free path of trapped neutrinos in compact cores of densities
about $10^{12} \, g.cm.^{-3}$ becomes smaller than the size of the stellar
core \cite{3,4}. Furthermore, the observational data collected from
supernovae 1987A indicates that the regime of radiation transport
prevailing during the emission process, is closer to the diffusion
approximation than to the streaming out limit \cite{5}.

However in many other circumstances, the mean free path of particles
transporting energy may be large enough as to justify the  free streaming
approximation. In this work, for simplicity, we shall consider  only the
streaming out limit.

As we shall see, in the relativistic regime, two (at least) different
definitions  of polytrope are
possible, yielding the same Newtonian limit. We shall consider both of
them, as possible ``seed'' equations of state and we shall contrast the
patterns of evolution obtained from each
case.

The plan of the paper is as follows. In Section 2 we define the conventions
and give the field equations and expressions for the kinematical and
physical variables we shall use, in
noncomoving coordinates. In Section 3 we present a brief review of the
properties of Newtonian polytropes and discuss two possible generalizations
to the relativistic regime. A resume
of the proposed approach is presented in Section 4. In Section 5 the method
is applied to the case when the ``seed'' equation  of state is a
relativistic polytrope and some examples are
explicitly worked out. Finally a discussion of results is presented in
Section 6.

\section{Relevant  Equations and Conventions}
\subsection{The field equations}
We consider spherically symmetric distributions of collapsing
fluid,
undergoing dissipation in the form of free streaming
radiation, bounded by a
spherical surface $\Sigma$.

\noindent
The line element is given in Schwarzschild--like coordinates by

\begin{equation}
ds^2=e^{\nu} dt^2 - e^{\lambda} dr^2 -
r^2 \left( d\theta^2 + sin^2\theta d\phi^2 \right),
\label{metric}
\end{equation}

\noindent
where $\nu(t,r)$ and $\lambda(t,r)$ are functions of their arguments. We
number the coordinates: $x^0=t; \, x^1=r; \, x^2=\theta; \, x^3=\phi$. We
use geometric units and therefore we have $c=G=1$.

\noindent
The metric (\ref{metric}) has to satisfy the Einstein field equations

\begin{equation}
G^\nu_\mu=-8\pi T^\nu_\mu,
\label{Efeq}
\end{equation}

\noindent
which in our case read \cite{Bo}:

\begin{equation}
-8\pi T^0_0=-\frac{1}{r^2}+e^{-\lambda}
\left(\frac{1}{r^2}-\frac{\lambda'}{r} \right),
\label{feq00}
\end{equation}

\begin{equation}
-8\pi T^1_1=-\frac{1}{r^2}+e^{-\lambda}
\left(\frac{1}{r^2}+\frac{\nu'}{r}\right),
\label{feq11}
\end{equation}

\begin{eqnarray}
-8\pi T^2_2  =  -  8\pi T^3_3 = & - &\frac{e^{-\nu}}{4}\left(2\ddot\lambda+
\dot\lambda(\dot\lambda-\dot\nu)\right) \nonumber \\
& + & \frac{e^{-\lambda}}{4}
\left(2\nu''+\nu'^2 -
\lambda'\nu' + 2\frac{\nu' - \lambda'}{r}\right),
\label{feq2233}
\end{eqnarray}

\begin{equation}
-8\pi T_{01}=-\frac{\dot\lambda}{r},
\label{feq01}
\end{equation}

\noindent
where dots and primes stand for partial differentiation with respect
to $t$ and $r$,
respectively.

\noindent
In order to give physical significance to the $T^{\mu}_{\nu}$ components
we apply the Bondi approach \cite{Bo}. Thus, following Bondi, let us
introduce purely locally Minkowski coordinates ($\tau, x, y, z$)
$$d\tau=e^{\nu/2}dt\,;\qquad\,dx=e^{\lambda/2}dr\,;\qquad\,
dy=rd\theta\,;\qquad\, dz=rsin\theta d\phi.$$

\noindent
Then, denoting the Minkowski components of the energy tensor by a bar,
we have
$$\bar T^0_0=T^0_0\,;\,\,
\bar T^1_1=T^1_1\,;\,\, \bar T^2_2=T^2_2\,;\,\,
\bar T^3_3=T^3_3\,;\,\, \bar T_{01}=e^{-(\nu+\lambda)/2}T_{01}.$$

\noindent
Next, we suppose that when viewed by an observer moving relative to these
coordinates with proper velocity $\omega$ in the radial direction, the physical
content  of space consists of a fluid of energy density $\rho$,
radial pressure $P$ and unpolarized radiation of energy density $\hat\epsilon$
traveling in the radial direction. Thus, when viewed by this (comoving with
the fluid)
observer the covariant tensor in
Minkowski coordinates is

\[ \left(\begin{array}{cccc}
\rho + \hat\epsilon    &  -\hat\epsilon  &   0     &   0    \\
-\hat\epsilon &  P + \hat\epsilon    &   0     &   0    \\
0       &   0       & P  &   0    \\
0       &   0       &   0     &   P
\end{array} \right). \]

\noindent
Then a Lorentz transformation readily shows that

\begin{equation}
T^0_0=\bar T^0_0= \frac{\rho + P \omega^2 }{1 - \omega^2} +
\epsilon,
\label{T00}
\end{equation}

\begin{equation}
T^1_1=\bar T^1_1=-\frac{ P + \rho \omega^2}{1 - \omega^2} -
\epsilon,
\label{T11}
\end{equation}

\begin{equation}
T^2_2=T^3_3=\bar T^2_2=\bar T^3_3=-P,
\label{T2233}
\end{equation}

\begin{equation}
T_{01}=e^{(\nu + \lambda)/2} \bar T_{01}=
-\frac{(\rho + P) \omega e^{(\nu + \lambda)/2}}{1 - \omega^2} -
e^{(\nu + \lambda)/2} \epsilon,
\label{T01}
\end{equation}
\noindent
with

\begin{equation}
\epsilon\equiv\hat\epsilon\frac{(1+\omega)}{(1-\omega)}.
\label{defepsilon}
\end{equation}

\noindent
Note that the coordinate velocity in the ($t,r,\theta,\phi$) system, $dr/dt$,
is related to $\omega$ by

\begin{equation}
\omega=\frac{dr}{dt}\,e^{(\lambda-\nu)/2}.
\label{omega}
\end{equation}

\noindent
Feeding back (\ref{T00}--\ref{T01}) into (\ref{feq00}--\ref{feq01}), we get
the field equations in  the form

\begin{equation}
\frac{\rho + P \omega^2 }{1 - \omega^2} +
\epsilon=-\frac{1}{8 \pi}\Biggl\{-\frac{1}{r^2}+e^{-\lambda}
\left(\frac{1}{r^2}-\frac{\lambda'}{r} \right)\Biggr\},
\label{fieq00}
\end{equation}

\begin{equation}
\frac{ P + \rho \omega^2}{1 - \omega^2} +\epsilon=-\frac{1}{8
\pi}\Biggl\{\frac{1}{r^2} - e^{-\lambda}
\left(\frac{1}{r^2}+\frac{\nu'}{r}\right)\Biggr\},
\label{fieq11}
\end{equation}

\begin{eqnarray}
P = -\frac{1}{8 \pi}\Biggl\{\frac{e^{-\nu}}{4}\left(2\ddot\lambda+
\dot\lambda(\dot\lambda-\dot\nu)\right)
 - \frac{e^{-\lambda}}{4}
\left(2\nu''+\nu'^2 -
\lambda'\nu' + 2\frac{\nu' - \lambda'}{r}\right)\Biggr\},
\label{fieq2233}
\end{eqnarray}

\begin{equation}
\frac{(\rho + P)}{(1 - \omega^2)}\, \omega e^{(\nu + \lambda)/2}
+e^{(\nu + \lambda)/2} \epsilon=-\frac{\dot\lambda}{8 \pi r}.
\label{fieq01}
\end{equation}

Observe that if $\nu$ and $\lambda$ are fully specified, then
(\ref{fieq00}--\ref{fieq01}) becomes a system of  algebraic equations for
the physical variables $\rho$, $P$,
$\omega$ and $\epsilon$.

\noindent
At the outside of the fluid distribution, the spacetime is that of Vaidya,
given by
\begin{equation}
ds^2= \left(1-2M(u)/\cal R\right) du^2 + 2dud{\cal R} -
{\cal R}^2 \left(d\theta^2 + sin^2\theta d\phi^2 \right),
\label{Vaidya}
\end{equation}
\noindent
where $u$ is a coordinate related to the retarded time, such that
$u=constant$ is (asymptotically) a
null cone open to the future and ${\cal R}$ is a null coordinate ($g_{{\cal
R}{\cal R}}=0$). It should
be remarked, however, that strictly speaking, the radiation can be considered
in radial free streaming only at radial infinity.

\noindent
The two coordinate systems ($t,r,\theta,\phi$) and ($u,{\cal
R},\theta,\phi$) are
related at the boundary surface and outside it by

\begin{equation}
u=t-r-2M\,ln \left(\frac{r}{2M}-1\right),
\label{u}
\end{equation}

\begin{equation}
{\cal R}=r.
\label{radial}
\end{equation}

\noindent
In order to match smoothly the two metrics above on the boundary surface
$r=r_\Sigma(t)$, we must require the continuity of the first and the second
fundamental
form across that surface.
Then it follows \cite{Herrera}

\begin{equation}
e^{\nu_\Sigma}=1-{2M}/{R_\Sigma},
\label{enusigma}
\end{equation}
\begin{equation}
e^{-\lambda_\Sigma}=1-{2M}/{R_\Sigma}.
\label{elambdasigma}
\end{equation}
\begin{equation}
\left[P\right]_\Sigma=0,
\label{PQ}
\end{equation}
where, from now on, subscript $\Sigma$ indicates that the quantity is
evaluated at the boundary surface $\Sigma$.
\noindent
Next, it will be useful to calculate the radial component of the
conservation law

\begin{equation}
T^\mu_{\nu;\mu}=0.
\label{dTmn}
\end{equation}
where
\begin{equation}
T_{\mu\nu} = \left(\rho+P\right)u_\mu u_\nu - P g_{\mu\nu} +
 \epsilon
l_\nu l_\mu
\label{T-}
\end{equation}
with
\begin{equation}
u^\mu=\left(\frac{e^{-\nu/2}}{\left(1-\omega^2\right)^{1/2}},\,
\frac{\omega\, e^{-\lambda/2}}{\left(1-\omega^2\right)^{1/2}},\,0,\,0\right),
\label{umu}
\end{equation}

\begin{equation}
l^\mu=\left(e^{-\nu/2},\,e^{-\lambda/2},\,0,\,0\right),
\label{null}
\end{equation}
where $u^\mu$ denotes the four velocity of the fluid,  and $l^\mu$ is a
null outgoing vector.

\noindent
After tedious but simple calculations we get

\begin{equation}
\left(-8\pi T^1_1\right)'=\frac{16\pi}{r} \left(T^1_1-T^2_2\right)
+ 4\pi \nu' \left(T^1_1-T^0_0\right) +
\frac{e^{-\nu}}{r} \left(\ddot\lambda + \frac{\dot\lambda^2}{2}
- \frac{\dot\lambda \dot\nu}{2}\right),
\label{T1p}
\end{equation}

\noindent
which in the static case becomes

\begin{equation}
P'=-\frac{\nu'}{2}\left(\rho+P\right),
\label{Prp}
\end{equation}

\noindent
which is the well known the Tolman--Oppenheimer--Volkoff equation.

\section{Newtonian and Relativistic Polytrope}
Although Newtonian polytropes  are well known and examined in detail in
most classical books on stellar structure \cite{Weigert}, 
we found it worthwhile to present here
the very basic facts about its theory.

\subsection{The Newtonian case}
As  mentioned before, the theory of polytropes is based on the assumption
of hydrostatic equlibrium, therefore the two starting equations are
(remember that we are using geometric
units)
\begin{equation}
\frac{dP}{dr}=-\frac{d\phi}{dr} \rho_{0},
\end{equation}
and
\begin{equation}
\frac{1}{r^2} \frac{d}{dr}(r^2\frac{d\phi}{dr})=4\pi \rho_{0},
\end{equation}
with $\phi$ and $\rho_{0}$  denoting the Newtonian gravitational potential
and the mass (baryonic) density, respectively.

Combining the two equation above with the polytropic equation of state
\begin{equation}
P=K\rho_{0}^{\gamma}=K\rho_0^{1+1/n} ,
\label{Pol}
\end{equation}
one obtains the well known Lane--Emden equation (for $\gamma\neq 1$)
\begin{equation}
\frac{d^2\psi_{0}}{d\xi^2}+\frac{2}{\xi}\frac{d\psi_{0}}{d\xi}+\psi_{0}^n=0
\end{equation}
with
\begin{equation}
r=\xi/A_{0},
\label{ere2}
\end{equation}

\begin{equation}
A_{0}^2=\frac{4 \pi \rho_{0c}^{(n-1)/n}}{K(n+1)},
\label{A2}
\end{equation}

\begin{equation}
\psi_{0}^n=\rho_{0}/\rho_{0c},
\label{psi1}
\end{equation}
where subscript $c$ indicates that the quantity is evaluated at the centre,
and the following boundary conditions apply:
$$\frac{d\psi_{0}}{d\xi}(\xi=0)=0;$$  $$\psi_{0}(\xi=0)=1.$$
The boundary surface of the sphere is defined by  $\xi=\xi_{n}$, such that
$\psi_{0}(\xi_{n})=0$.

As it is well known, bounded configurations exist only for $n<5$ and
analytical solution may be found for $n=0, 1$ and $5$.

It is also worth remembering that the polytropic equation of state may be
used to model two different type of situations, namely:
\begin{itemize}

\item When the polytropic constant $K$ is fixed and can be calculated from
natural constants. This is the case of a completely degenerate gas in the
non--relativistic ($\gamma=5/3$; $n=3/2$) and relativistic limit
($\gamma=4/3$; $n=3$).
\item
When $K$ is a free parameter as for example in the case of isothermal ideal
gas or in a completely convective star.
\end{itemize}

\subsection{The relativistic case}
When considering the polytropic equation of state within the context of
general relativity, two distinct expressions are often considered. In order
to avoid confussion we shall
differentiate them from the begining. Thus, the following two cases may be
contemplated.

\subsection{Case I}
In this case the original polytropic equation of state is conserved
\begin{equation}
P=K \rho_0^{1+1/n} ,
\label{PolI}
\end{equation}
then it follows from the first law of thermodynamics that
\begin{equation}
d(\frac{\rho + P}{\cal N})-\frac{dP}{\cal N}=Td(\frac{\sigma}{\cal N}) ,
\label{1ley}
\end{equation}
where $T$ denotes temperature, $\sigma$ is entropy per  unit of
proper volume and $\cal N$ is the particle density, such that
\begin{equation}
\rho_0={\cal N} m_0.
\label{n}
\end{equation}
Then for an adiabatic process it follows
\begin{equation}
d(\frac{\rho}{\cal N})+Pd(\frac{1}{\cal N})=0 ,
\label{1ley11}
\end{equation}
which together with (\ref{PolI}) leads to
\begin{equation}
K\rho_{0}^{\gamma-2}=\frac{d(\rho/\rho_{0})}{d\rho_{0}} ,
\label{Poladiabdif}
\end{equation}
with
\begin{equation}
\gamma=1+1/n .
\label{gamma}
\end{equation}
If $\gamma\neq 1$, (\ref{Poladiabdif}) may be easily integrated to give
\begin{equation}
	\rho= C \rho_0 +  P/(\gamma-1).
\label{densnueva}
\end{equation}

In the non--relativistic limit we should have $\rho\rightarrow\rho_{0}$, and
therefore $C=1$.
Thus, the polytropic equation of state amounts to
\begin{equation}
	\rho= \rho_0 +  P/(\gamma-1).
\label{densnuevaI}
\end{equation}
It is worth noticing that the familiar ``barotropic'' equation of state
\begin{equation}
	\rho=  P/(\gamma-1) ,
\label{densnuevaII}
\end{equation}
is a particular case of (\ref{densnueva}) with $C=0$.

In the very special case $\gamma=1$, one obtains
\begin{equation}
K\rho_{0}^{-1}=\frac{d(\rho/\rho_{0})}{d\rho_{0}} ,
\label{Poladiabdif3}
\end{equation}
whose solution is
\begin{equation}
	\rho= P\log\rho_{0} +\rho_{0} C,
\label{densgamma1}
\end{equation}
or, if puting $C=1$ from the non--relativistic limit
\begin{equation}
	\rho= P\log\rho_{0} +\rho_{0}.
\label{densgamma11}
\end{equation}
From now on we shall only consider the $\gamma\neq 1$ case.

Next, let us introduce the following variables
\begin{equation}
\alpha=P_c/\rho_{c},
\label{alfa}
\end{equation}
\begin{equation}
r=\xi/A,
\label{ere}
\end{equation}
\begin{equation}
A^2=4 \pi \rho_{c}/[\alpha (n+1)],
\label{A}
\end{equation}
\begin{equation}
\psi_{0}^n=\rho_{0}/\rho_{0c},
\label{psi}
\end{equation}
\begin{equation}
v(\xi)=m(r) A^3/(4 \pi\rho_{c}),
\label{ve}
\end{equation}
where the mass function, as usually is defined by
\begin{equation}
e^{-\lambda}=1-2m/r.
\label{mass}
\end{equation}
Then the Tolman--Oppenheimer--Volkoff equation (\ref{Prp}) becomes
\begin{equation}
\xi^2 \frac{d\psi_{0}}{d\xi}(\frac{1-2(n+1)\alpha
v/\xi}{(1-\alpha)+(n+1)\alpha \psi_{0}})+v+\alpha\xi^3 \psi_{0}^{n+1}=0,
\label{TOV1}
\end{equation}
and from the definition of mass function and equation (13) in the static
case, we have
\begin{equation}
m'=4 \pi r^2 \rho
\label{mprima}
\end{equation}
or
\begin{equation}
\frac{dv}{d\xi}=\xi^2 \psi_{0}^n (1-n\alpha+n\alpha\psi_{0}).
\label{veprima}
\end{equation}
In the Newtonian limit ($\alpha\rightarrow0$), (\ref{TOV1}) and
(\ref{veprima}) become
\begin{equation}
\xi^2 \frac{d\psi_{0}}{d\xi}+v=0
\label{TOV1Newt}
\end{equation}
and
\begin{equation}
\frac{dv}{d\xi}=\xi^2 \psi_{0}^n,
\label{veprimaNewt}
\end{equation}
which are equivalent to the classical Lane--Emden equation
\begin{equation}
\frac{d^2\psi_{0}}{d\xi^2}+\frac{2}{\xi}\frac{d\psi_{0}}{d\xi}+\psi_{0}^n=0.
\end{equation}

\subsection{Case II}
Sometimes  it is  assumed that the relativistic polytrope is defined by
\begin{equation}
P=K \rho^{1+1/n},
\label{PolII}
\end{equation}
instead of (\ref{PolI}).
Then introducing
\begin{equation}
\psi^n=\rho/\rho_{c},
\label{psi2}
\end{equation}
related to $\psi_{0}$ by
\begin{equation}
\psi^n=\psi_{0}^{n}(1-n\alpha+\alpha n\psi_{0}).
\label{psi2n}
\end{equation}
The  TOV equation becomes
\begin{equation}
\xi^2 \frac{d\psi}{d\xi}(\frac{1-2(n+1)\alpha v/\xi}{1+\alpha
\psi})+v+\alpha\xi^3 \psi^{n+1}=0,
\label{TOV2}
\end{equation}
and from (\ref{mprima})
\begin{equation}
\frac{dv}{d\xi}=\xi^2 \psi^n.
\label{veprima2}
\end{equation}
In the Newtonian limit   ($\alpha\rightarrow0$), the Lane--Emden equation
is also
recovered in this case, as it should be.

\begin{figure}
\centerline{\epsfxsize=4.5in\epsfbox{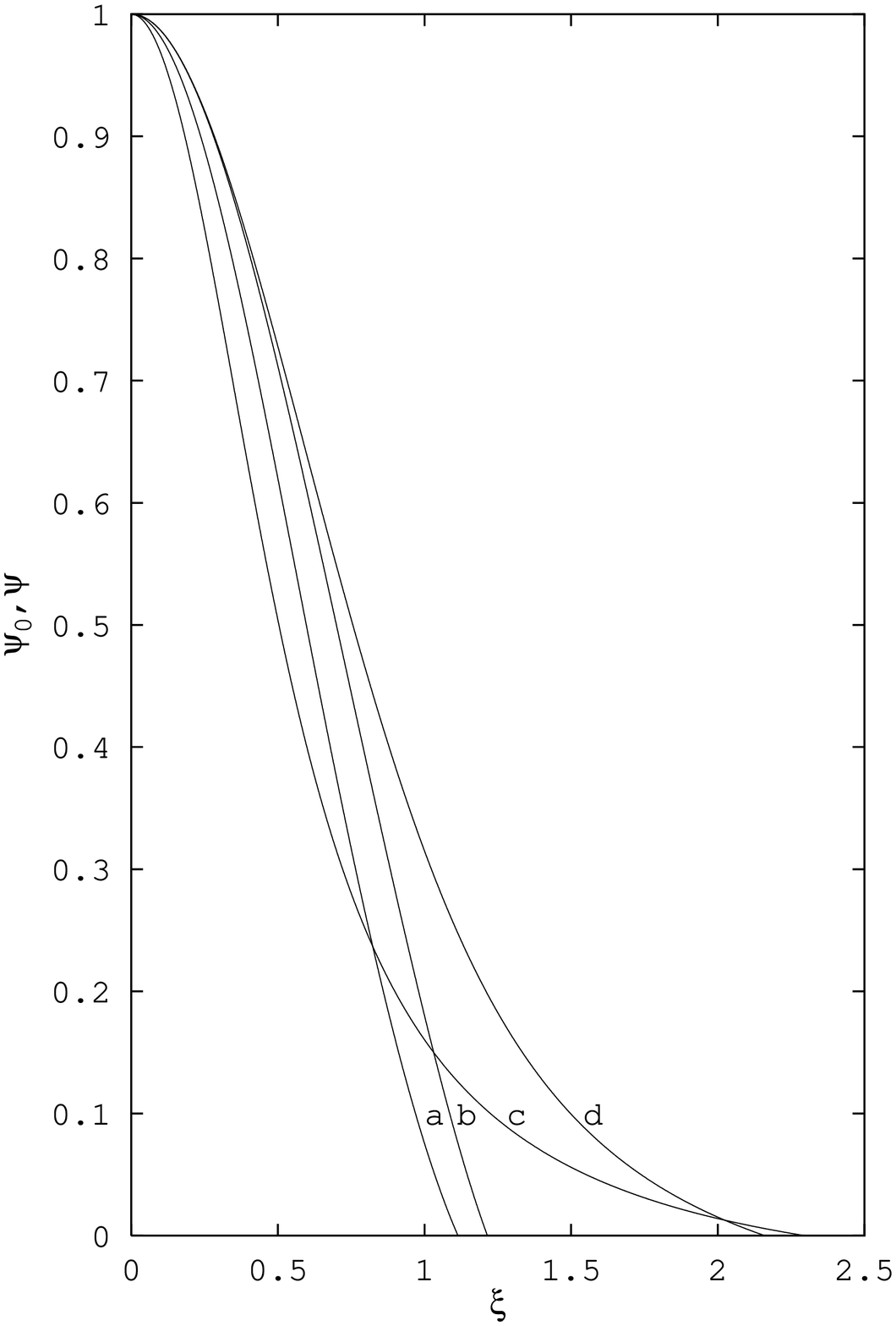}}
\caption{$\psi_0$ (Case I) and $\psi$ (Case II) as a function of
$\xi$ for $\alpha=1$ and different
values of $n$: (a) Case I, $n=0.5$; (b) Case II, $n=0.5$; (c) Case I, $n=1.5$;
(d) Case II, $n=1.5$.}
\label{fig:psiI_II_1}
\end{figure}

\begin{figure}
\centerline{\epsfxsize=4.5in\epsfbox{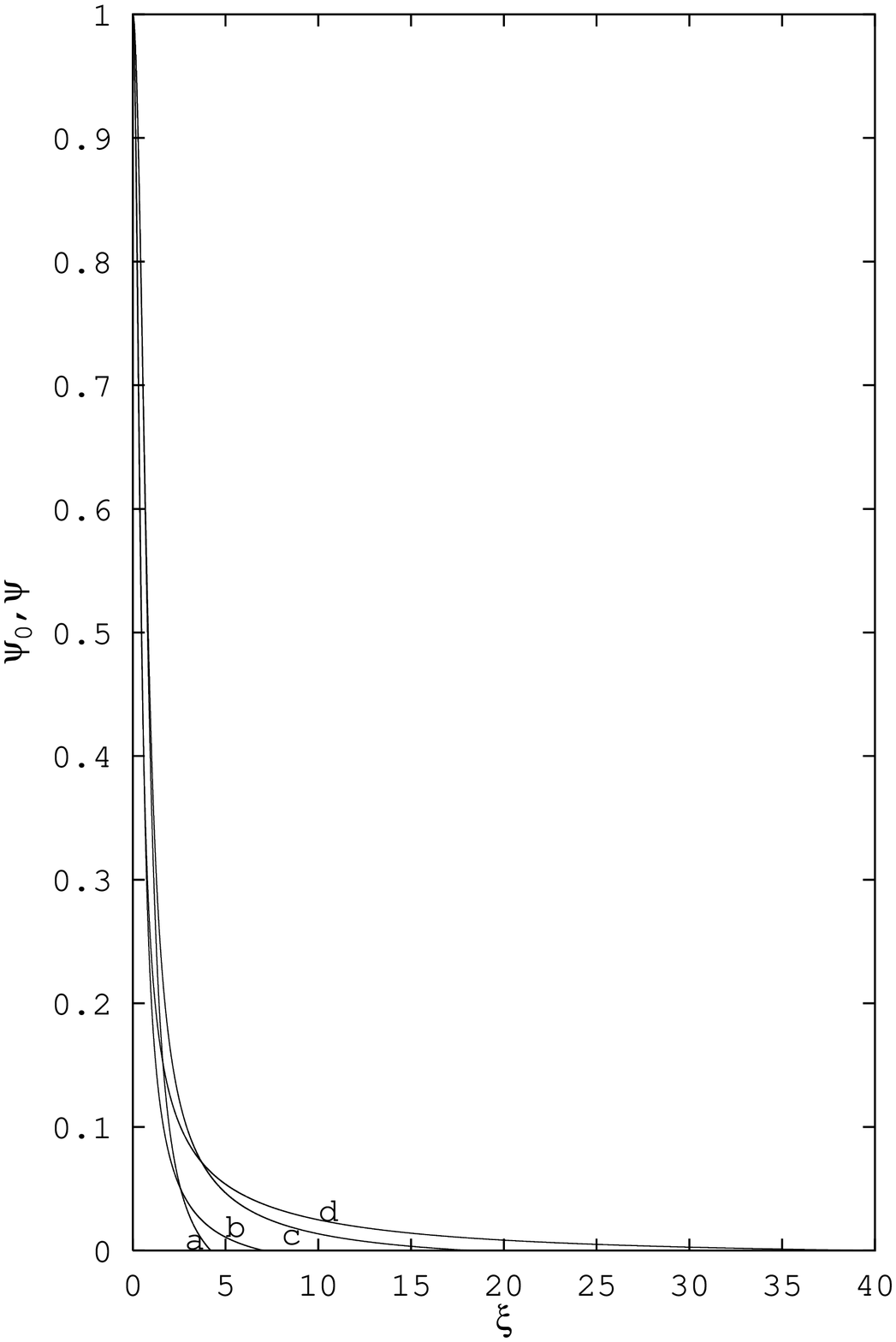}}
\caption{$\psi_0$ (Case I) and $\psi$ (Case II) as a function of
$\xi$ for $\alpha=1$ and different
values of $n$: (a) Case II, $n=2$; (b) Case I, $n=2$; (c) Case II, $n=2.5$;
(d) Case I, $n=2.5$.}
\label{fig:psiI_II_2}
\end{figure}

Obviously both equations of state differ each other, specially in the
highly relativistic regime. This can be verified by inspection of figures
\ref{fig:psiI_II_1} and \ref{fig:psiI_II_2}.

\section{The method}

Let us now give a brief resume of the method we shall use to describe the
evolution of the relativistic polytrope.
However before doing so some general considerations will be necessary.

\subsection{Equilibrium and quasi--equilibrium}
The simplest situation, when dealing with self--gravitating spheres, is that
of equilibrium (static case). In our notation that means that
$\omega=\epsilon=0$,  all time derivatives
vanishes, and we obtain the generalized Tolman--Oppenheimer--Volkoff equation
(\ref{Prp}).

Next, we have the quasistatic regime. By this we mean that the sphere
changes slowly, on a time scale that is very long compared to the typical
time in which the sphere reacts to a slight perturbation of hydrostatic
equilibrium, this typical time scale is called hydrostatic time scale
\cite{Weigert} (sometimes this time scale is also referred to as dynamical
time scale, e.g. see
\cite{Kawaler}). Thus, in this regime  the system is always very close to
hydrostatic
equilibrium and its evolution may be regarded as a sequence of static
models linked by (\ref{fieq01}). This assumption is very sensible because
the hydrostatic
time scale is very small for many phases of the life of the star.
It is of the order of $27$ minutes for the Sun, $4.5$ seconds for a white dwarf
and $10^{-4}$ seconds for a neutron star of one solar mass and $10$ Km radius.
It is well known that any of the stellar configurations mentioned above,
generally,
change on a time scale that is very long compared to their respective
hydrostatic time scales.

However, as already mentioned, in some important cases, this approximation
is not longer reliable, and one needs to consider departures from
quasi--equilibrium. We shall describe such
departures, in the post--quasi--static approximation defined below.

\subsection{The effective variables and the post--quasistatic approximation}

Let us now define the following effective variables:

\begin{equation}
\tilde\rho=T^0_0= \frac{\rho + P\omega^2 }{1 - \omega^2} +
 \epsilon,
\label{rhoeffec}
\end{equation}

\begin{equation}
\tilde P=-T^1_1=\frac{ P + \rho \omega^2}{1 - \omega^2} +\epsilon.
\label{peffec}
\end{equation}

 In the quasistatic regime the effective variables satisfy the same
equation (\ref{Prp}) as the corresponding physical variables (taking into
account the contribution of $\epsilon$ to the ``total'' energy density and
radial pressure, whenever
the free streaming approximation is being used). Therefore in
the quasistatic situation (and obviously
 in  the static too), effective and physical variables share the same
radial dependence.
Next, feeding back (\ref{rhoeffec}) and (\ref{peffec}) into (\ref{fieq00})
and (\ref{fieq11}), these two equations may be formally integrated, to
obtain:

\begin{equation}
m = 4 \pi \int^{r}_{0}{r^2 \tilde\rho dr}, \\
\label{m}
\end{equation}

\begin{equation}
\nu = \nu_\Sigma + \int^{r}_{r_\Sigma}\frac{2 (4 \pi r^3  \tilde
P+m)}{r(r-2m)} dr.
\label{nu}
\end{equation}
From where it is obvious that for a given radial dependence of the
effective variables, the radial dependence of metric functions becomes
completely determined.

With this last comment in mind, we shall define the post--quasistatic regime
as that corresponding to  a system out of equilibrium (or quasiequilibrium)
but whose effective variables
share the same radial dependence as the corrresponding physical variables
in the state of equilibrium (or quasiequilibrium). Alternatively it may be
said that the system in the
post--quasistatic regime is characterized by metric functions whose radial
dependence is the same as the metric functions corresponding to the static
(quasistatic) regime. The rationale
behind this definition is not difficult to grasp: we look for a regime
which although out of equilibrium, represents the closest possible
situation to a quasistatic evolution (see more
on this point in the last Section).

\subsection{The algorithm}

Let us now outline the approach that we shall use:
\begin{enumerate}
\item  Take an interior solution to Einstein equations, representing a fluid
distribution of matter in equilibrium, with a given
$$\rho_{st}=\rho(r);\,\qquad\, P_{st}= P(r).$$
This static solution will be obtained in  this work by integration of the
relativistic Lane--Emden equations (\ref{TOV1}), (\ref{veprima}) or
(\ref{TOV2}), (\ref{veprima2}).
\item  Assume that the $r$ dependence of $\tilde P$ and $\tilde\rho$ is the
same as that of $P_{st}$ and $\rho_{st}$, respectively.

\item  Using equations  (\ref{m}) and (\ref{nu}), with the $r$ dependence of
$\tilde P$ and $\tilde\rho$, one gets $m$ and $\nu$ up to some functions of
$t$, which will be specified below.

\item  For these functions of $t$ one has three ordinary differential equations
(hereafter referred to as surface equations), namely:
\begin{enumerate}
\item  Equation (\ref{omega}) evaluated  on $r=r_{\Sigma}$.

\item The equation relating the total mass loss rate with the energy flux
through the boundary surface.

\item  Equation (\ref{T1p}) evaluated on $r=r_{\Sigma}$.
\end{enumerate}
\item The system of
surface equations described above may be closed with the additional
information about some of the physical variables evaluated on
the boundary surface (e.g. the
luminosity).

\item Once the system of surface equations is closed, it may be integrated for
any particular initial data.

\item  Feeding back the result of integration in the expressions for $m$ and
$\nu$, these two functions are completely determined.

\item  With the input from the point 7 above, and remembering that once
metric functions are fully specified, field equations become an algebraic
system of equations for the
physical variables; these may be found for any piece of
matter distribution.
\end{enumerate}
\subsection{The Surface equations}

 As it should be clear from the above, the crucial point in the algorithm is
the system of surface equations. So, let us specify them now.

Introducing the dimenssionles variables
$$A=r_{\Sigma}/m_{\Sigma}(0),$$
$$F=1-2M/A,$$
$$M=m_{\Sigma}/m_{\Sigma}(0),$$
$$\Omega=\omega_{\Sigma},$$
$$\beta=t/m_{\Sigma}(0),$$
where  $m_{\Sigma}(0)$ denotes the total initial mass, we obtain  the first
surface equation by
evaluating (\ref{omega}) at $r=r_{\Sigma}$. Thus, one gets
\begin{equation}
\frac{dA}{d\beta}=F\Omega. \label{eq:first}
\end{equation}

Next, using junction conditions, one obtains from (\ref{mass}),
(\ref{fieq00}) and (\ref{fieq01})
evaluated at $r=r_{\Sigma}$, that
\be
\frac{dM}{d\beta}=-F (1+\Omega)\hat E,
\label{flux}
\ee
with
\be
\hat E=4 \pi r^{2}_{\Sigma}\hat \epsilon_{\Sigma},
\ee
where the first and second term on the right of (\ref{flux}) represent the
gravitational redshift and the Doppler shift corrections, respectively.

Then, defining the luminosity perceived by an observer at infinity as
$$L=-\frac{dM}{d\beta}.$$
we obtain the second surface equation in the form
\be
\frac{dF}{d\beta}=\frac FA(1-F)\Omega +2 L/A. \label{eq:second}
\ee

The third surface equation may be obtained by evaluating at the boundary
surface the conservation law
$T_{1;\mu }^\mu=0$, which reads
\begin{eqnarray}
\tilde P^{'} + \frac{(\tilde\rho + \tilde P)(4\pi r^3\tilde P + m)}{r(r-2m)}
=\nonumber \\ \frac{e^{-\nu}}{4\pi r(r-2m)}\left( \ddot m +\frac{3\dot
m^2}{r-2m}-
\frac{\dot m \dot \nu}{2}\right) + \frac{2}{r}(P-\tilde P). \label{eq:TOV}
\end{eqnarray}

In the case when the effective density is separable, i.e.,
$\tilde\rho={\cal F}(t){\cal H}(r)$;
equation (\ref{eq:TOV}) evaluated at the boundary surface leads to
\begin{eqnarray}
\frac{d\Omega}{d\beta}=\Omega^2\left[\frac{8F}{A}+2F{\cal
K}(r_{\Sigma})+4\pi\tilde
\rho_{\Sigma} A(3-\Omega^2)\right] \nonumber \\
-\frac{F}{\tilde\rho_{\Sigma}}\left[R+\frac{2}{A}\left(\tilde\rho_{\Sigma}
\Omega^2
+\frac{\bar E(1+\Omega)}{4\pi r_{\Sigma}^2}\right)\right],
\label{eq:TOV_a}
\end{eqnarray}
where
\begin{equation}
R=\left[\tilde P^{'} + \frac{\tilde P+\tilde\rho}{1-2m/r}(4\pi r\tilde P
+\frac{m}{r^2})\right]_{\Sigma},
\label{R}
\end{equation}

\be
\bar E=\hat E (1+\Omega)
\ee
and
\begin{equation}
{\cal K}(r_{\Sigma})=\frac{d
}{dr_{\Sigma}}ln\left(\frac{1}{r_{\Sigma}}\int^{r_{\Sigma}}_0 dr r^2
{\cal H}(r)/{\cal H}(r_{\Sigma})\right).
\end{equation}

Before analyzing specific models, some interesting conclusions can be
obtained at this level of
generality.
One of these conclusions concerns the condition of bouncing
at the surface which, of course, is related to the occurrence of a minimum
radius $A$. According to (\ref{eq:first}) this requires $\Omega=0$, and we have
\begin{equation}
\frac{d^{2} A}{d{\beta^2}} = F\frac{d \Omega}{d \beta},
\end{equation}
or using (\ref{eq:TOV_a})
\begin{equation}
\frac{d\Omega}{d\beta}(\Omega=0)=-\frac{F}{\tilde\rho_{\Sigma}}\left[R+\frac
{2 \hat E}{4\pi r_{\Sigma}^{2} A}\right]. \label{eq:TOV_a0}
\end{equation}
Observe that a positive  energy flux ($\hat E$)
tends to decrease the radius of the
sphere, i.e., it favors the compactification of the object, which is easily
understood. The same happens
when $R>0$. The opposite effect occurs when these
quantities have the opposite
signs. Now, for a positive energy flux the sphere
can only bounce at its surface when
$$\frac{d\Omega}{d\beta}(\Omega=0) \ge 0.$$
According to (\ref{eq:TOV_a0})
this requires
\begin{equation}
-R(\Omega=0)\ge 0.
\label{bounce}
\end{equation}

A physical meaning can be associated to this equation as follows. For
non--radiating, static configuration, $R$ as defined by (\ref{R}) consists
of two parts.
The first term which
represents the hydrodynamical force (see (\ref{Prp})) and the second which
is of course the gravitational force. The
resulting force in the sense of increasing $r$ is precisely $-R$, if this
is positive a net outward
acceleration occurs and vice--versa.
Equation (\ref{bounce}) is the natural generalization of this result for
general non--static configurations.

\section{Models and their numerical implementation}
\subsection{Effective variables}
Once the profiles of energy density and pressure have been established in
the satic case via
the integration of the corresponding Lane--Emden equations, we proceed with
the determination of effective
variables according to the algorithm sketched above. However, as it should
be clear such determination is not unique. The following possibilities
arise:

\begin{enumerate}
\item $\tilde\rho=f(t)+h(r)$
$\tilde p=g(t)+i(r)$,
where $h(r)$ and $i(r)$ correspond to the pressure and total energy density
obtained from the integration of the relativistic  Lane--Emden equations,
in both cases described above. However this model  has not static limit.

\item $\tilde\rho=f(t)h(r)$ and $\tilde p=g(t) + K\tilde\rho_0^{1+1/n}$,
for the case I, where $\tilde\rho_0=f(t)h_0(r)$, being $h_0(r)$
the baryonic mass density in the static limit;
 $\tilde p=g(t) + K\tilde\rho^{1+1/n}$, for the case  II. In both cases
$K=K(m_{\Sigma},r_{\Sigma})$.
\end{enumerate}

On what follows we shall consider only the possibility 1 above.

\subsection{ Numerical implementation of models}

We have used an standard Runge--Kutta routine to obtain functions $h(r)$,
$h_0(r)$ and
$i(r)$ from the integration of relativistic   Lane--Emden equations
for different values of  $n$ and $\alpha$. Integration was performed from
$\xi=0$ until the first zero of $\psi$ (or $\psi_{0}$).

Next, for the third surface equation we need to calculate numerically the
following terms:
\begin{equation}
\left[\frac{d i(r)}{d
r}\right]_{r=r_{\Sigma}},
\label{diidr}
\end{equation}
\begin{equation}
k(t)=\int_0^{r_{\Sigma}} r^2 h(r)
dr.\label{intache}\end{equation}

Observe that  $dk(t)/dt=0$, since $h(r_{\Sigma})=0$.

For the calculation of  (\ref{diidr}) and
(\ref{intache}) we have adjusted a  Chebyshev polynomial \cite{Che} to
functions  $h(r)$ and $i(r)$. Also, for the calculation of either of these
functions in the  Chebyshev's nodes or
within different interior regions we have used the interpolating
Lagrange polynomials.

A standard  Runge--Kutta method has also been applied to solve surface
equations.
These three equations are solved as an initial value problem, upon
specification of $A(t=0)$, $F(t=0)$ and one function of $u$. Specifically
we choose
$$L=\frac{2 m_{R}}{\sqrt{\pi}}\, e^{-4(t-5/2)^2},$$
where $m_R$ is the mass to be radiated.

Once the surface equations have been integrated, we proceed to calculate
the metric functions and their derivatives. For doing so, we need to
calculate numerically the following
expressions:
\begin{equation}\frac{di(r)}{dr},\label{t1}\end{equation}
\begin{equation}\int^r_0 r^2 h(r) dr,\label{t2}\end{equation}
\begin{equation}\int^r_a \frac{2(4\pi r^3 \tilde p + m)}{r(r-2m)}dr\label
{t3}\end{equation}
and
\begin{equation}\int^r_a \frac{\partial}{\partial t}
\left\{ \frac{2(4\pi r^3 \tilde p +
m)}{r(r-2m)}\right\}dr.\label{t4}\end{equation}
Where the last expression appears in the equation for the time derivative
of $\nu$ given by
$$\dot \nu=\dot\nu_{\Sigma} + \int^r_{r_{\Sigma}(t)} \frac{\partial}{\partial t}
\left\{ \frac{2(4\pi r^3 \tilde p + m)}{r(r-2m)}\right\}dr
-\left\{ \frac{2(4\pi r^3 \tilde p + m)}{r(r-2m)}\right\}_{r_{\Sigma}(t)}
\dot r_{\Sigma}$$
For the numerical integration of (\ref{t3}) and (\ref{t4})
it is necessary to calculate the integrands on points of the latice defined in
the integration of Lane--Emden equation, using again  the
Chebyshev's polynomials and
Lagrange interpolants.
Once the metric functions and their derivatives have been completely
determined, we use the field equations to obtain algebraically the physical
variables.

All along evolution we keep radial dependence obtained from the solution of
the Lane--Emden
equations. This
was implemented fitting
the $h(r)$ and $i(r)$ profiles to the radius's distribution at time $t$.
Thus, the radial coordinate is scaled by means of:
$$r\rightarrow\left(\frac{r_{\Sigma}(t)}{r_{\Sigma}(0)}\right)r.$$

The developed code was paralelized using MPI routines for FORTRAN.
We use as many nodes as interior regions studied. One tipical run
takes, for one region and one time unit, one a half hour in a
900 MHz. central processing unit.

\subsection{Models}
Although a large number of models has been worked out, we shall present
here only two for illustration, corresponding to the cases I and II. Both
were calculated for
values: $n=2$, $\alpha=0.1$, $\Omega(0)=-0.05$, with an emission of
$0.01$ of the initial mass. The profiles of physical variables are
exhibited in figures 3--6. As we increase the emission we arrive at a
point where case II becomes unphysical before
case I. If we increase $n$, for both cases, the initial distribution is less
compact. On the contrary, if we increase $\alpha$ the initial distribution
is more compact. Figure 7 shows the normalized radii evolution for both
cases, different
values of $n$ and $\alpha=0.1$.

\section{Conclusions}
We have considered two possible definitions of relativistic polytrope and
have presented a method to study their evolution. The models represent a
generalization of the static
polytrope to the case of evolving and dissipating fluid spheres, which in
the static limit satisfy a polytropic equation of state. This allows for
modeling self--gravitating
objects, and at the same time brings out differences between the two
possible definitions of polytropes, considered here. We have incorporated
dissipation, a fundamental process in
the process of gravitational collapse, into discussion. It remains, to
consider the diffusion limit, however because of the additional
complication associated to the necessity of
introducing an equation of transport, we have only considered here the
simplest, streaming out limit.

Although the  examples are presented with the sole purpose of illustrating
the method (our main goal here being to provide a tool for modeling the
evolution of relativistic
polytropes), some comments on them, are in order.

 The difference between the two definitions of polytrope considered here,
are clearly exhibited in figures 1--2. To magnify such difeference we
present the results corresponding to
the ``ultra--relativistic'' case  ($\alpha=1$). As can be seen, for $n\le
1.365$, configurations of case I have smaller radii than those
corresponding to the case II. This situation
reverses for $n>1.365$. In general, bounded relativistic configurations
exist for smaller values of $n$ , than in the Newtonian case.

Figures 3--6 show how differently, both polytropes evolve. As it appears
from these figures, the case II leads to an stronger collpase. This
tendency is confirmed by curves $c-d$
of figure 7. Also, curves $a-b$ on this same figure show an example of
bouncing for $n=2.5$. The strongest bouncing of case I, further indicates
that the equation of state
resulting from case I is stiffer than the obtained from case II. It is
worth mentioning that these differences are observed in a large number of
models, for a wide range of values of
$n$, $\alpha$ and initial data.

\section*{Acknowledgements}
WB was benefited from research support from
FONACIT under grant S1--98003270. Computer time for this work was provided by the
Centro Nacional de C\'alculo Cient\'\i fico de la Universidad
de los Andes (CeCalcULA).

\begin{figure}
\centerline{\epsfxsize=3in\epsfbox{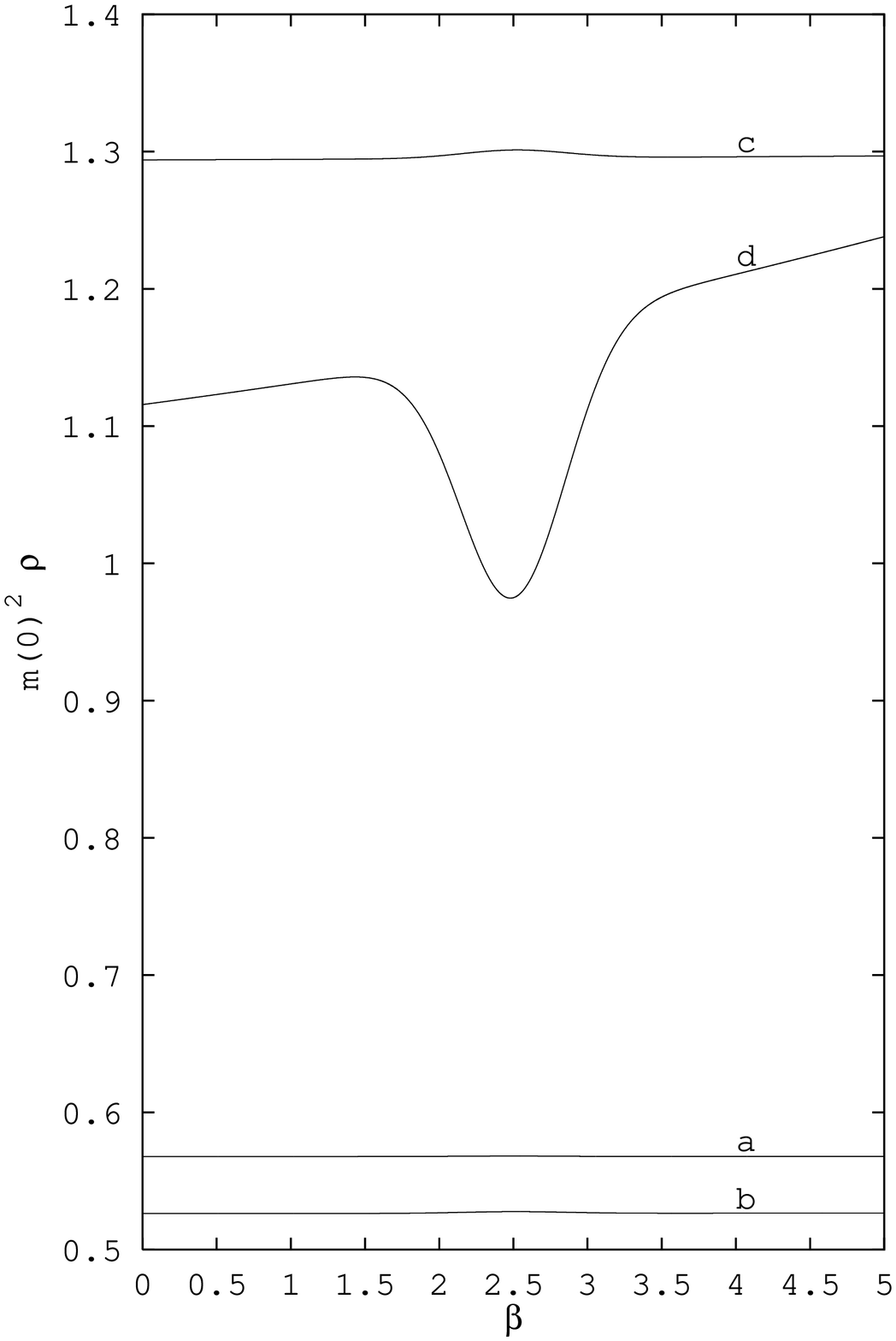}\epsfxsize=3in\epsfbox{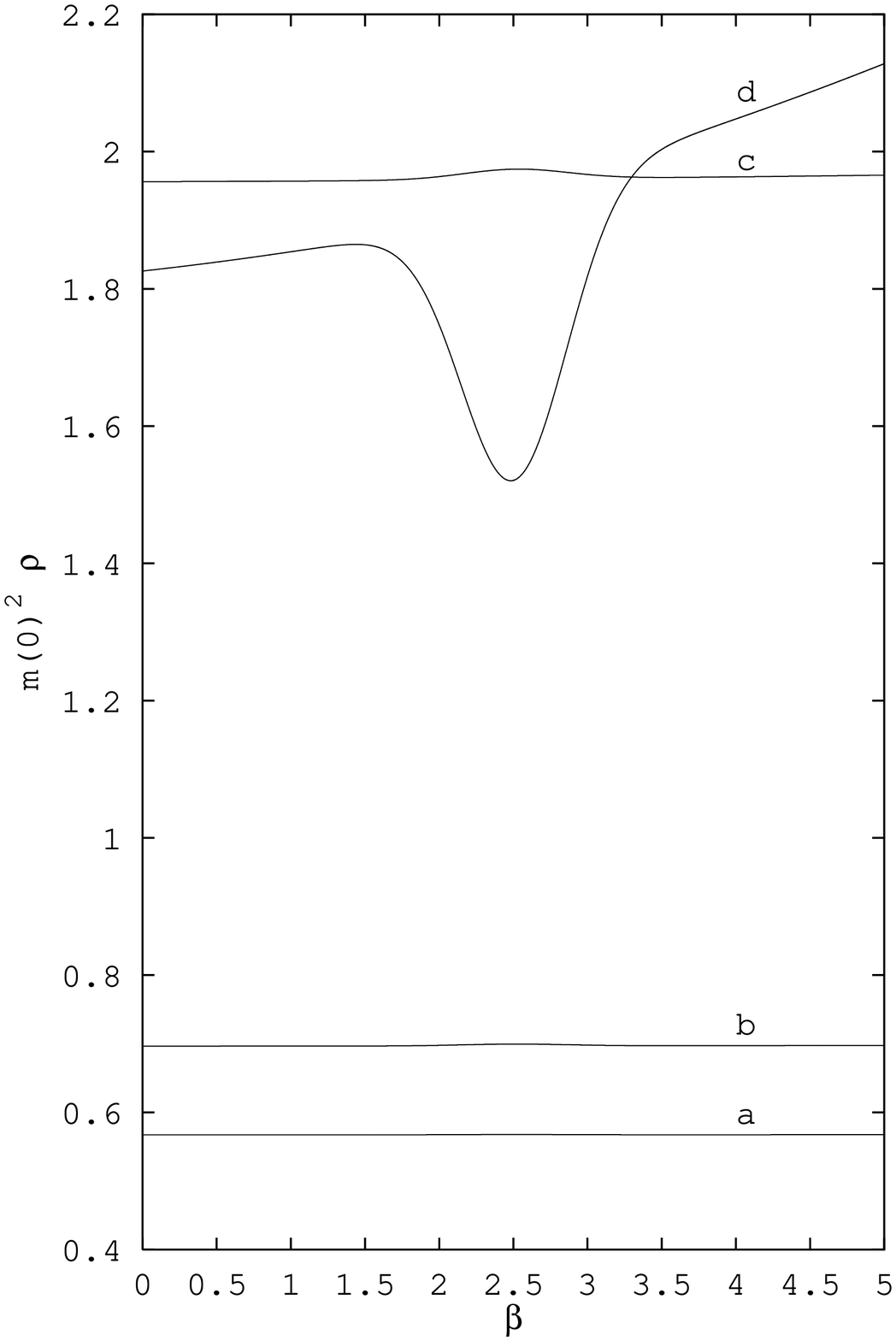}}
\caption{Adimensional energy density (Case I to the left; Case II to the right)
 as a function of dimenssionles time for $n=2$
and $\alpha=0.1$ at different regions: (a) $r/a=0.25$ (multiplied by 10);
(b) $r/a=0.50$ (multiplied by 10); (c) $r/a=0.75$ (multiplied by $10^2$);
(d) $r/a=1.00$ (multiplied by $10^4$).}
\label{fig:dens}
\end{figure}
\begin{figure}
\centerline{\epsfxsize=3in\epsfbox{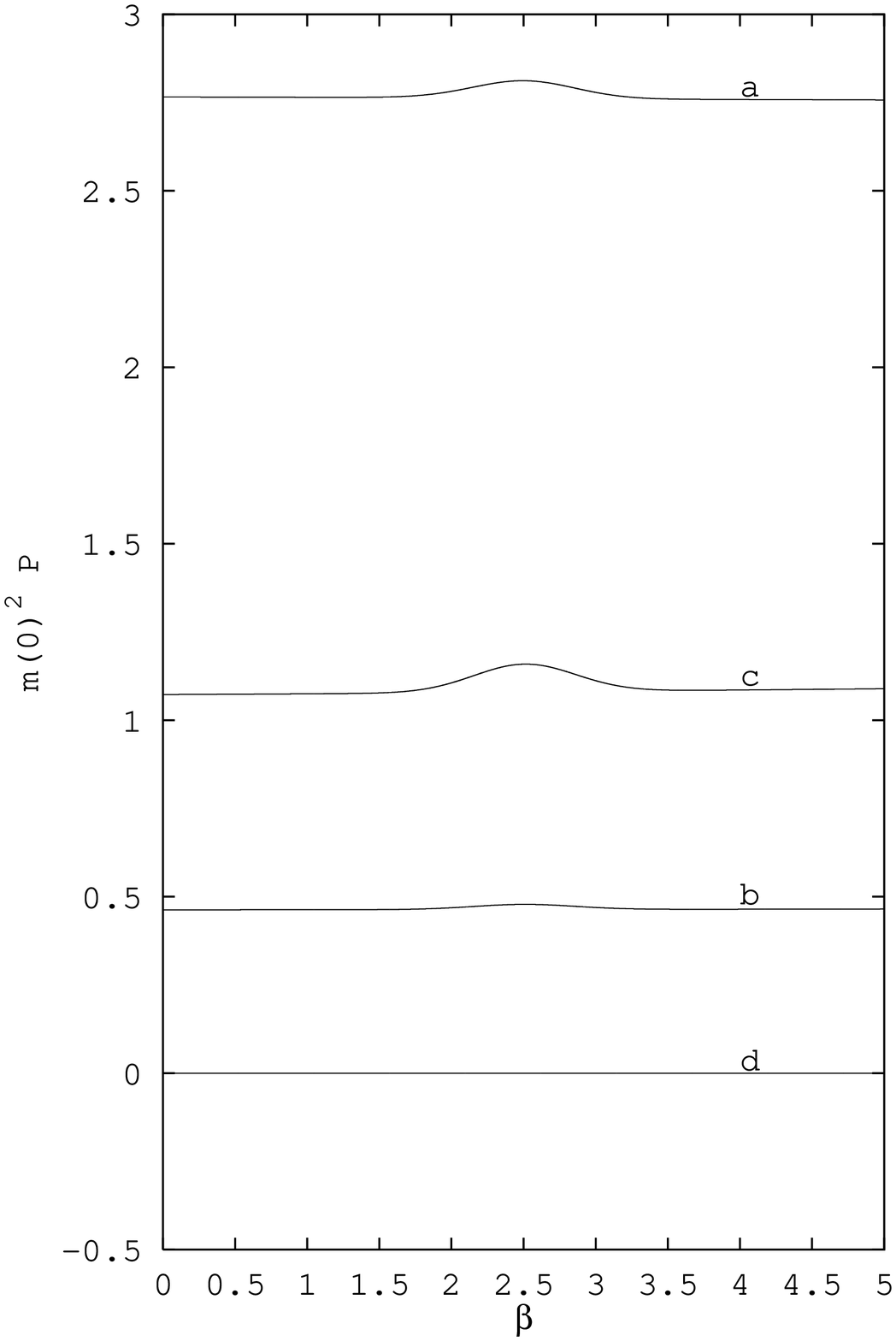}\epsfxsize=3in\epsfbox{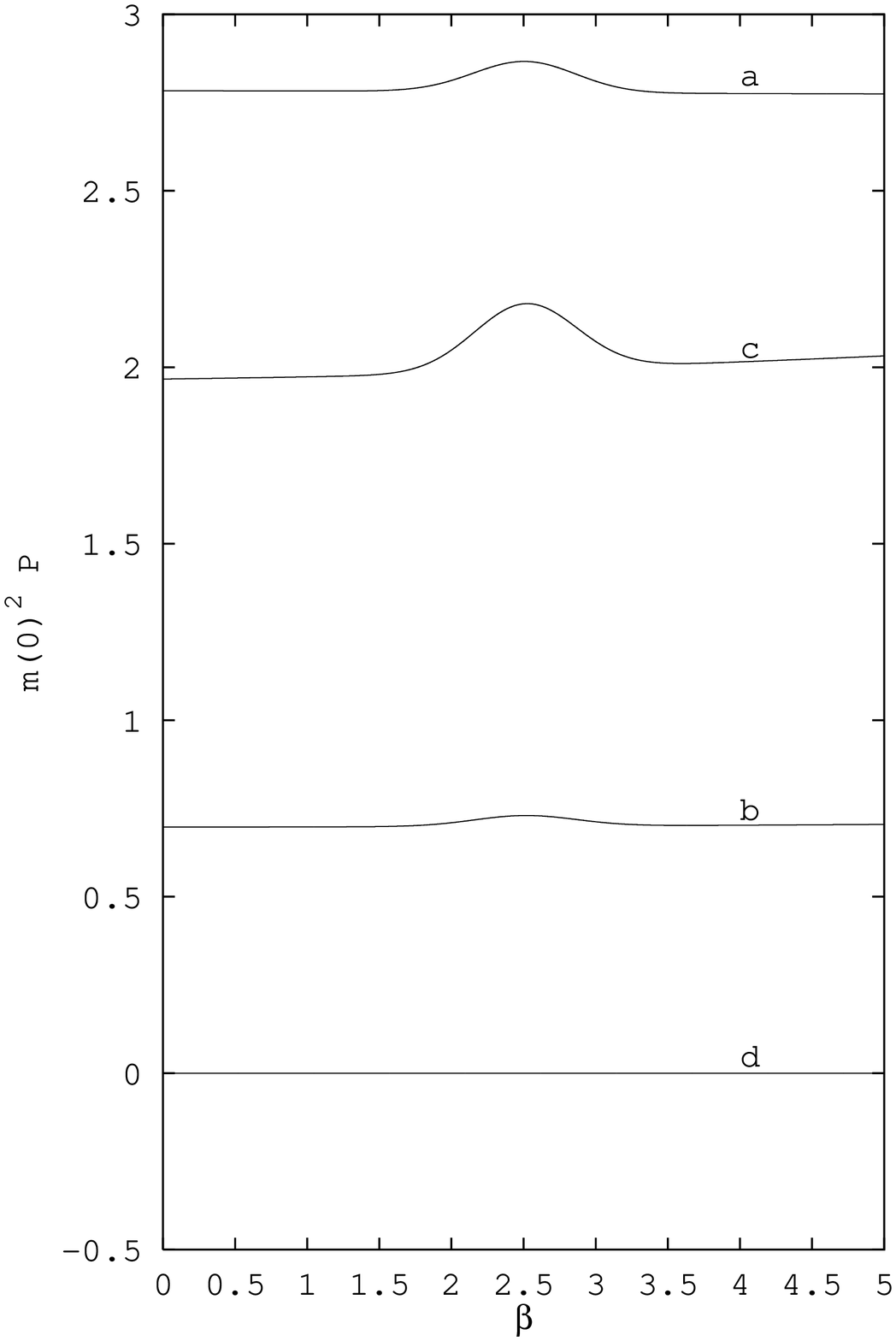}}
\caption{Adimensional pressure (Case I to the left; Case II to the right)
as a function of dimenssionles time for
$n=2$ and $\alpha=0.1$ at
different regions:
(a) $r/a=0.25$ (multiplied by $10^3$); (b) $r/a=0.50$ (multiplied by
$10^2$); (c) $r/a=0.75$ (multiplied by $10^3$); (d) $r/a=1.00$.}
\label{fig:press}
\end{figure}
\begin{figure}
\centerline{\epsfxsize=3in\epsfbox{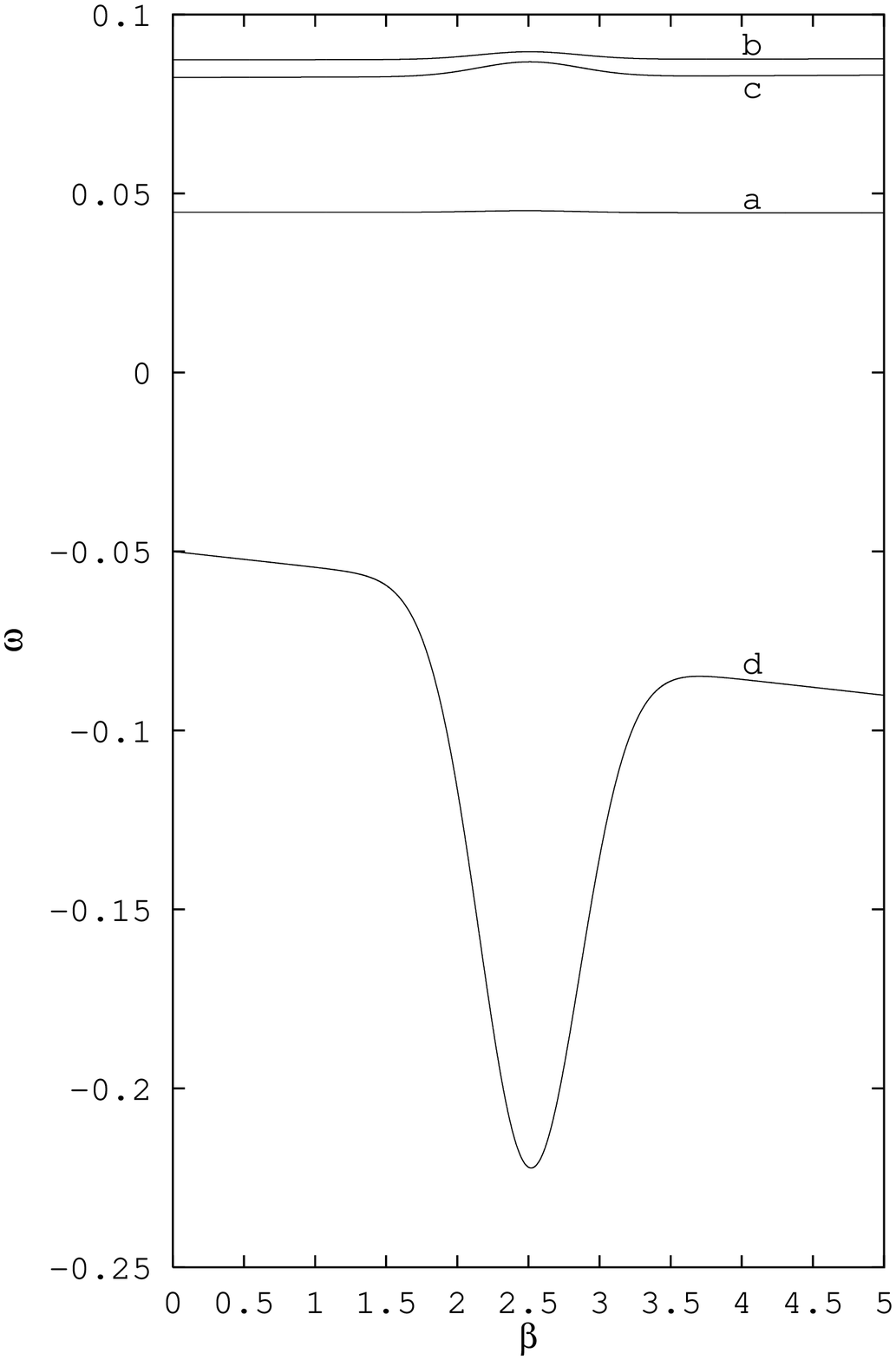}\epsfxsize=3in\epsfbox{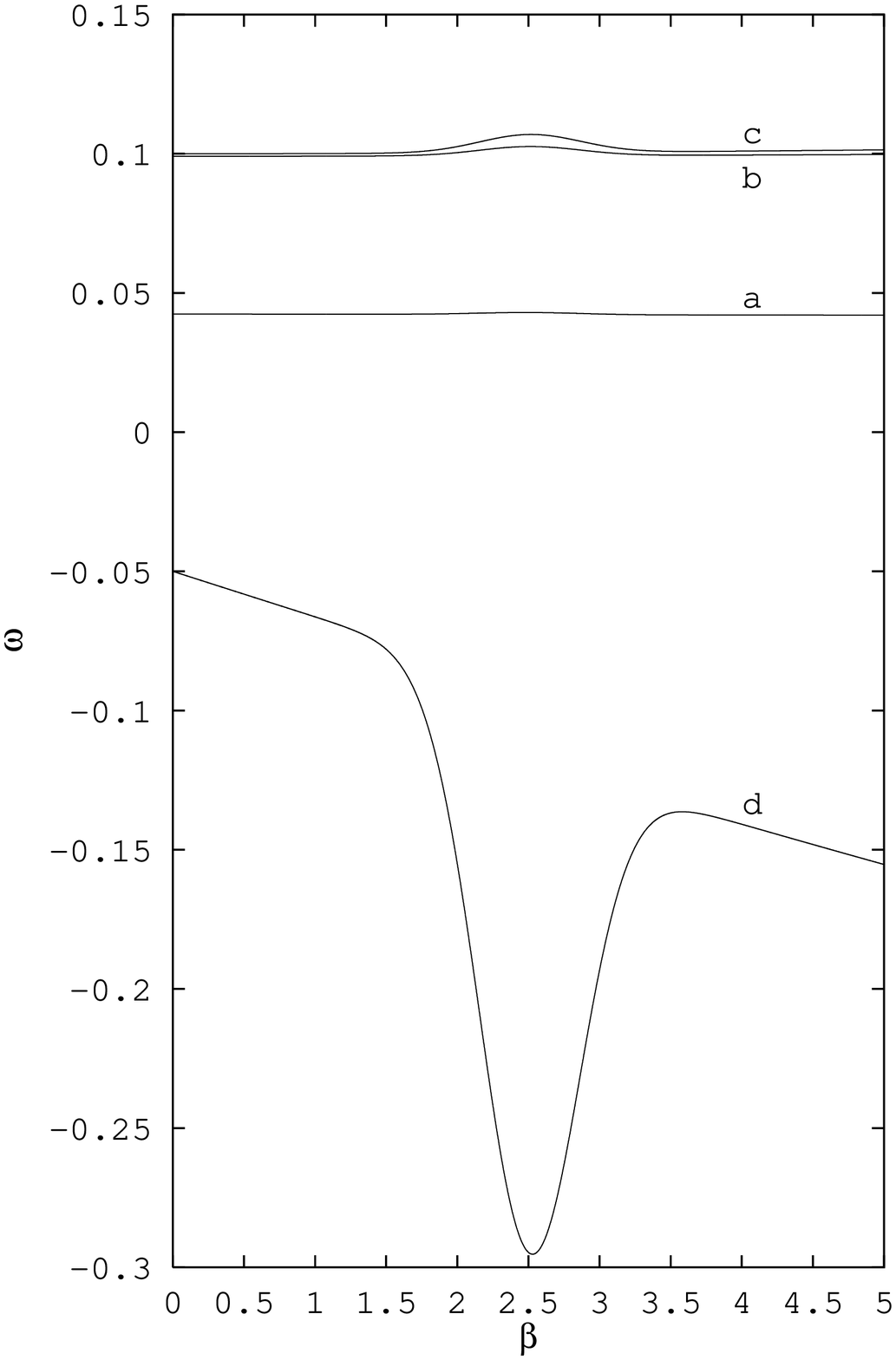}}
\caption{Radial velocity (Case I to the left; Case II to the right)
as a function of dimenssionles time for $n=2$ and $\alpha=0.1$ at
different regions:
(a) $r/a=0.25$; (b) $r/a=0.50$; (c) $r/a=0.75$; (d) $r/a=1.00$.}
\label{fig:velo}
\end{figure}
\begin{figure}
\centerline{\epsfxsize=3in\epsfbox{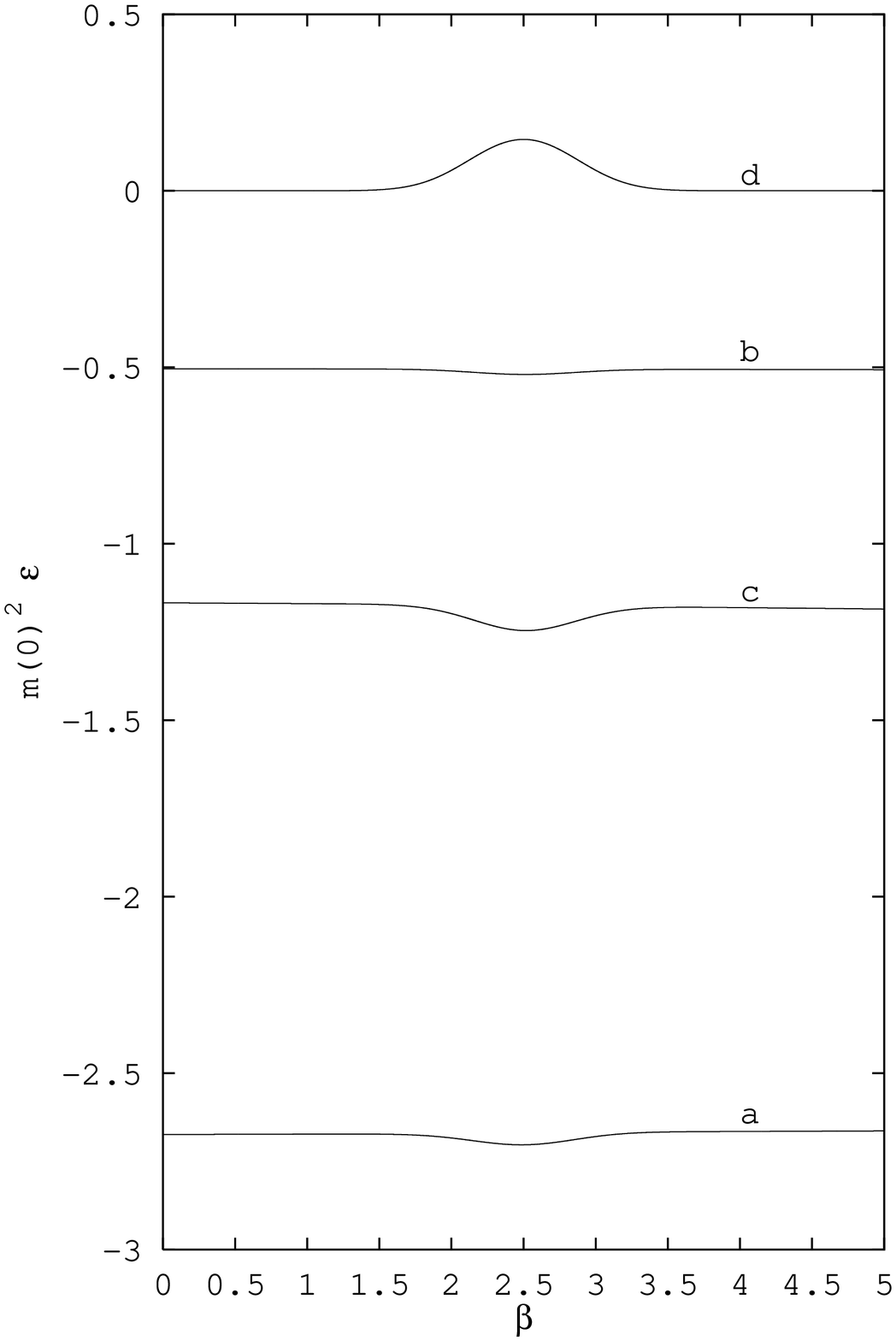}\epsfxsize=3in\epsfbox{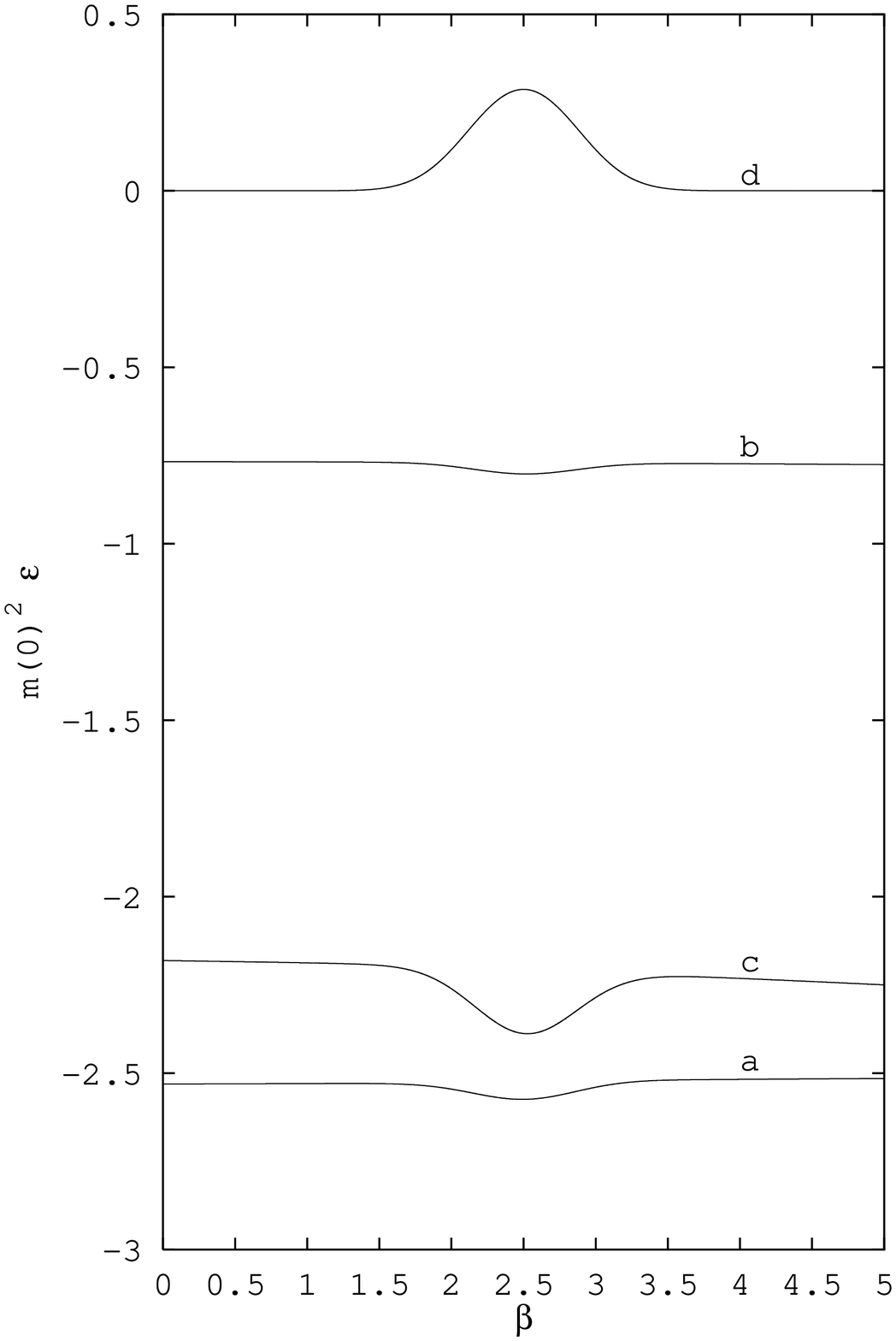}}
\caption{Adimensional flux of energy (Case I to the left; Case II to the right)
as a function of dimenssionles time for
$n=2$ and $\alpha=0.1$ at
different regions:
(a) $r/a=0.25$ (multiplied by $10^3$); (b) $r/a=0.50$ (multiplied by $10^2$);
(c) $r/a=0.75$ (multiplied by $10^3$); (d) $r/a=1.00$ (multiplied by $10^4$).}
\label{fig:flux}
\end{figure}
\begin{figure}
\centerline{\epsfxsize=4.5in\epsfbox{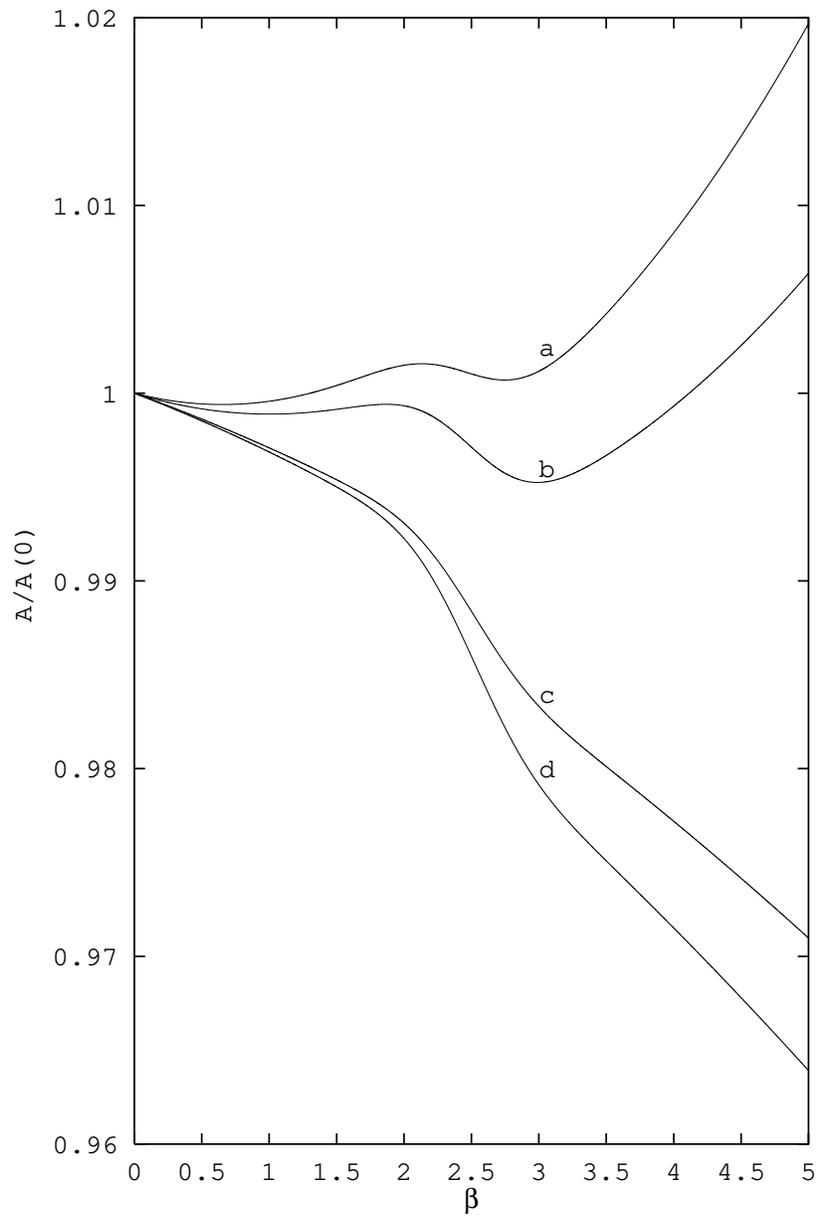}}
\caption{Evolution of the normalized radii for $\alpha=0.1$, both cases and
different
values of $n$: (a) Case I, $n=2.5$; (b) Case II, $n=2.5$; (c) Case I,
$n=1.5$; (d) Case II, $n=1.5$. }
\label{fig:radii}
\end{figure}

\end{document}